\documentclass[10pt,letterpaper]{article}
\usepackage{opex3}
\usepackage{color}
\usepackage{cite}
\usepackage{float}
\begin{document}

\title{Translative lens-based \\full field coherent X-ray imaging}

\author{Carsten Detlefs$^1$,Mario A. Beltran$^2$, Jean-Pierre Guigay$^1$, and Hugh Simons$^{2*}$}

\address{$^{1}$ European Synchrotron Radiation Facility, B.P. 220, F-38043 Grenoble Cedex, France \\
$^2$ Physics Department, Technical University of Denmark, 2800 Kgs. Lyngby, Denmark}
\email{$^*$	husimo@fysik.dtu.dk} 



\begin{abstract}
We describe a full-field coherent imaging approach suitable for hard X-rays  based on a classical (i.e. Galilean) X-ray microscope. The method combines a series of low-resolution images acquired at different transverse lens positions into a single high-resolution image, overcoming the spatial resolution limit set by the numerical aperture of the objective lens. We describe the optical principles of the approach, demonstrate the successful reconstruction of simulated phantom data, and discuss aspects of the reconstruction. We believe this approach offers some potential benefits over conventional scanning X-ray ptychography in terms of acquisition speed, spatial bandwidth and radiation dose rate.
\end{abstract} 


\section{Introduction}

Lens-based full-field X-ray microscopy, in which an objective lens between the object and detector creates a magnified image of the object, offers the possibility to image extended objects in a single acquisition. As such, it is well-suited for investigating dynamic processes, such as in materials \cite{Snigireva2018}, chemical reactions \cite{Meirer2011} and biological systems \cite{Werner2001}. However, the spatial resolution of a lens-based full-field microscope is physically limited by the finite numerical aperture (NA) of its objective lens, which tends to be small (0.01 or less) at hard X-ray energies (E $>$ 15 keV). Recent developments in X-ray optics have yielded substantial improvements in NA \cite{Morgan2015,Schroer2005}, but often at the cost of reducing the working distance to an impractical degree.

Synthetic aperture microscopy offers an alternative route to increasing the NA. One such approach is Fourier Ptychographic Microscopy (FPM) \cite{Zheng2013}, which involves combining a series of low-resolution intensity images in Fourier space and subsequently back-propagating to the object plane to recover the exit surface complex wavefield. Varying the angle of the incident full-field illumination samples a wider range of scattering directions, thus improving the space-bandwidth product without the need to move the sample, objective lens or detector \cite{Lohmann1996}. FPM's image recovery procedure therefore differs from that of conventional X-ray ptychography (for example, see Ref.~\cite{Rodenburg1992, Faulkner2004,Rodenburg2004, Rodenburg2007, Thibault2008,Maiden2009, Dierolf2010, Maiden2010, Humphry2012} $et$ $al$) in that the object support constraints are imposed in Fourier space rather than real space. The original implementation of FPM used a conventional optical microscope (i.e visible light) with a small magnification ($\times$2 objective) and NA (0.08) to achieve a synthetic NA of 0.5, resulting in a spatial resolution comparable to a $\times$20 objective while maintaining the much larger field of view and depth of field of the original low-magnification configuration. 

Adapting FPM to the X-ray regime could potentially address two key shortcomings of lens-based full-field X-ray microscopy: The compound image corresponds to a larger, synthetic NA, while digital wavefront correction may be used during the image recovery procedure to compensate for lens aberrations (which may be appreciable)  \cite{Koch2016}. Furthermore, as this recovery procedure yields a complex image, one could exploit the phase contrast to dramatically increase sensitivity to weakly-interacting objects \cite{Cloetens1997}. We foresee that a practical X-ray implementation of FPM will require subtle differences to the original approach, however, since: (1) Rotating the incident beam is (as described in \cite{Zheng2013}) impractical/impossible at large-scale facilities (e.g. synchrotrons). This means one must instead rotate the entire optical axis of the system, likely causing alignment and image registration issues; and (2) for a grand variety of specimens, the scattering due to X-ray radiation is weak, with refraction angles often not exceeding the microradian range. To address these drawbacks, we alternatively propose moving the position of the lens \emph{transversely} to the optical axis and collecting images at various overlapping regions. This approach is, in principle, similar to pinhole-scanning methods (e.g. \cite{tsai2016,Faulkner2004,guizar-sicairos2008}), albeit using a focusing lens instead of a pinhole. We believe this offers some practical advantages.

In this paper, the theory and methodology outlining the idea of lens translation imaging (LTI) is structured as follows. Section~\ref{ForwardProblem} describes the image formation problem via mathematical formalisms pertinent to scalar coherent wavefield propagation. The LTI image acquisition method is depicted in section~\ref{Lens translation} and accompanied with numerical simulation examples. In section~\ref{InverseProblem} we detail the iterative phase-retrieval process that reconstructs the wavefield of the exit surface of the imaged object (see fig.~\ref{Fig:FourierPtychoSetup}). Results from numerical simulations are also shown.

\section{Theory of image formation (forward problem)} \label{ForwardProblem}

\begin{figure}[H]
\centering
\includegraphics[width=130mm]{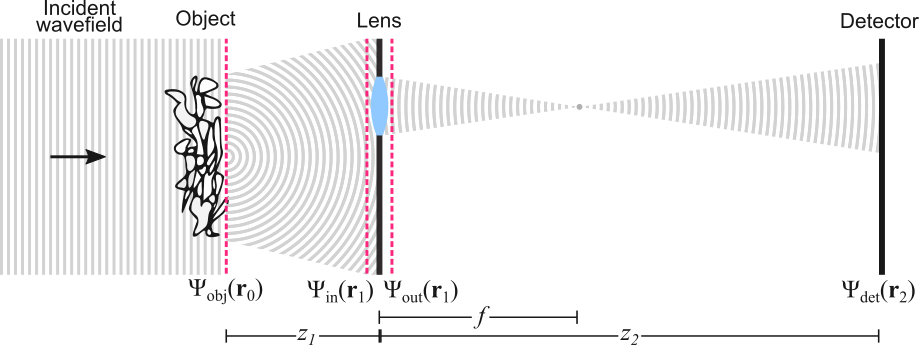}
\caption{Schematic of the lens translation imaging (LTI) setup.} 
\label{Fig:FourierPtychoSetup}
\end{figure}

Figure~\ref{Fig:FourierPtychoSetup} illustrates the LTI imaging configuration. Monochromatic coherent X-ray plane-wave electromagnetic illumination with wavelength $\lambda$ propagates rightwards along optical axis $z$. The incident radiation traverses an object or specimen where the intensity and phase changes incurred are imprinted on the complex wavefield $\Psi_{obj}$ exiting the object. This exit field then propagates downstream a distance $z_{1}$ reaching the entry plane $\Psi_{in}$ of an optically thin converging lens with finite aperture size and focal length $f$. The field transmitted through the lens $\Psi_{out}$ propagates a further distance $z_{2}$ to give $\Psi_{det}$, where a spatially sensitive detector is placed that measures the square modulus (intensity) of the wavefield $\left | \Psi_{det} \right |^{2}$, thus excluding all phase information.    

\noindent We derive an analytical expression for the wavefield $\Psi_{det}$ utilizing the linear operator theory of imaging \cite{Nazar1980}. The formalism treats the propagation and passage of optical wavefields though a system as a linear operator acting on some input to yield an associated output. This enables the problem in Fig.~\ref{Fig:FourierPtychoSetup} to be undertaken in a ``cascading'' approach, resulting in the following expression for the wavefield at the detector plane:         

\begin{eqnarray} 
\Psi_{det} (\textbf{r}_{2}) = \mathcal{P}_{z_{2}} \left [ T(\textbf{r}_{1}-\textbf{s}_{n})\mathcal{P}_{z_{1}}\left \{ \Psi_{obj} (\textbf{r}_{0}) \right \} \right ].   
\label{Eqn:FieldAtDetector}
\end{eqnarray}

\noindent Here, $\textbf{r}_{0}=(x_{0},y_{0})$, $\textbf{r}_{1}=(x_{1},y_{1})$ and $\textbf{r}_{2}=(x_{2},y_{2})$ are the  Cartesian coordinates normal to $z$ corresponding to the object ($\Psi_{obj}$), lens ($\Psi_{in}$ and $\Psi_{out}$) and detector plane ($\Psi_{det}$), respectively. $ T(\textbf{r}_{1}-\textbf{s}_{n})$ is the transmission function of the lens. 

The operator $\mathcal{P}_{z}$, which forward propagates a complex wavefield by a distance $z$, is the operator form of the Fresnel diffraction integral \cite{Goodman2005,Paganin2006,BornWolf}:

\begin{eqnarray} 
\Psi_{z} (\textbf{r}_{j+1}) &=& \mathcal{P}_{z}\left \{ \Psi (\textbf{r}_{j}) \right \}  \nonumber\\
&=& \frac{-i }{ \lambda z}\exp\left ( \frac{i2 \pi z}{\lambda } \right )\exp\left ( \frac{i \pi \left | \textbf{r}_{j+1}  \right |^{2}}{\lambda z} \right ) \mathcal{F}_{\textbf{r}_{j+1}}  \left \{  \exp\left ( \frac{i \pi \left | \textbf{r}_{j}  \right |^{2}}{\lambda z} \right ) \Psi (\textbf{r}_{j})  \right \}.
\label{Eqn:QPF}
\end{eqnarray}

\noindent This operator maps a field in a plane defined by the coordinates $\textbf{r}_{j}=(x_{j},y_{j})$ onto a propagated field defined by the coordinates $\textbf{r}_{j+1}=(x_{j+1},y_{j+1})$, where $j$ takes on non-negative integer values ($j=0,1$). For example, $j=0$ would correspond to the field mapping of $\Psi (\textbf{r}_{0})\rightarrow \Psi (\textbf{r}_{1})$. The operation acts from right to left as follows: (i) multiply the input wavefield by a quadratic phase factor in $\textbf{r}_{j}$; (ii) take the ``scaled'' Fourier transform which projects a complex function from $\textbf{r}_{j}$ to $\textbf{r}_{j+1}$; (iii) multiply the result by quadratic phase factor in $\textbf{r}_{j+1}$ and constant complex phase shift set by $z$. The Fourier transform convention used here is:

\begin{eqnarray} 
\mathcal{F}_{\textbf{r}_{j+1}}  \left \{ g (\textbf{r}_{j})\right \}&=&\int \int_{-\infty }^{\infty}g (\textbf{r}_{j})\exp\left ( -\frac{2 \pi i }{z\lambda} \textbf{r}_{j+1} \cdot \textbf{r}_{j}  \right ) d\textbf{r}_{j}    \nonumber\\
&=&  \int \int_{-\infty }^{\infty}g (\textbf{r}_{j})\exp\left ( -2 \pi i \textbf{k}_{j}\cdot \textbf{r}_{j}  \right ) d\textbf{r}_{j}     \nonumber\\
\textup{where},& &\textbf{k}_{j}=\frac{\textbf{r}_{j+1}}{z\lambda}.
\label{Eqn:FTconvention}
\end{eqnarray}

\noindent We note the use of the term ``scaled'', as the Fourier transform used here differs slightly from a conventional Fourier transform in that it maps a complex function from real space onto another complex function also in real space that is re-sampled by a scale factor $1/ \lambda z$ \cite{Goodman2005,Paganin2006}. 

The complex transmission function of the lens denoted by $ T(\textbf{r}_{1}-\textbf{s}_{n})$ can be decomposed into four separate functions corresponding to the phase shift $ Q_{f}^{\textbf{r}_{1}}$, absorption $A_{\textbf{s}_{n}}^{\textbf{r}_{1}}$, lens aberration $\chi(\textbf{r}_{1},\alpha_{pq})$ and masking $H_{\textbf{s}_{n}}^{\textbf{r}_{1}}$ (due to the finite aperture size) of the transmitted wavefield. That is:       

\begin{eqnarray} 
&T(\textbf{r}_{1}-\textbf{s}_{n})=H_{\textbf{s}_{n}}^{\textbf{r}_{1}} A_{\textbf{s}_{n}}^{\textbf{r}_{1}}Q_{f}^{\textbf{r}_{1}} \exp\left [ i\chi(\textbf{r}_{1},\alpha_{pq})  \right ], & \nonumber\\
& \textup{where},\: A_{\textbf{s}_{n}}^{\textbf{r}_{1}}=\exp \left [- \frac{\left | \textbf{r}_{1}- \textbf{s}_{n} \right |^{2}}{2 \sigma^{2}} \right ] ,\:   Q_{f}^{\textbf{r}_{1}}=\exp \left [- \frac{i \pi \left | \textbf{r}_{1}- \textbf{s}_{n} \right |^{2}}{\lambda f} \right ]. &
\label{Eqn:LensFunction}
\end{eqnarray}

\noindent The vector $\textbf{s}_{n}=(s_{x},s_{y})$ with magnitude $\left |\textbf{s}_{n}  \right |$ represents the translation position of the centre of the lens in relation to the $\textbf{r}_{1}$ plane. The index $n$ is an integer used as an indicator of the lens position. The complex function $Q_{f}^{\textbf{r}_{1}}$ quantifies the phase shifts imparted by the, assumed to be thin lens, on the entering wavefield $\Psi_{in}$ \cite{Paganin2006}. These phase shifts are consistent with the condition needed to create focus fields, where the phase exiting the surface of the lens must be such that a spherical wave is collapsed towards a point \cite{Paganin2006}. The amplitude attenuation suffered by $\Psi_{in}$ is determined by the function $A_{\textbf{s}_{n}}^{\textbf{r}_{1}}$, which has Gaussian profile characterized by the variance $\sigma$. $H_{\textbf{s}_{n}}^{\textbf{r}_{1}}$ represents the finite aperture size of the lens and serves to transmit only the spatial frequencies of the wavefield $\Psi_{in}$ within the radius of the physical aperture \cite{Simons2017}. $\chi(\textbf{r}_{1},\alpha_{pq})$ is the aberration function of the lens. The function is characterized by the \emph{aberration coefficients} $\alpha_{pq}$, where $p$ and $q$ are non-negative integers representing the aberration order \cite{BornWolf}. In the case of an ideal thin lens, as is considered in this study, we assume zero aberrations are present ($\chi(\textbf{r}_{1},\alpha_{pq})=0$). Note however that in  practical settings, these lens aberrations need to be either corrected using aberration balancing techniques or iteratively refined in the phase retrieval process - similar to the determination of the illumination function in classical ptychography. This however, is beyond the scope of this work.

Given that all terms and symbols have been defined in operator form, Eq.~\ref{Eqn:FieldAtDetector} can now be expressed as:     

\begin{eqnarray} 
 \Psi_{det} (\textbf{r}_{2}) &=& C_{z_{2}}^{z_{1}}Q_{z_{2},f}^ {\textbf{r}_{2},\textbf{s}_{n}}
\mathcal{F}_{\textbf{r}_{2}}  \left \{H_{\textbf{s}_{n}}^{\textbf{r}_{1}}A_{\textbf{s}_{l}}^{\textbf{r}_{1}}Q_{z_{1},z_{2}}^{\textbf{r}_{1},f} \exp \left ( \frac{-2 \pi i}{\lambda f}\textbf{s}_{n}\cdot\textbf{r}_{1}\right ) \mathcal{F}_{\textbf{r}_{1}}\left [Q_{z_{1}}^{\textbf{r}_{0}} \Psi_{obj} (\textbf{r}_{0}) \right ] \right \}.  \nonumber\\  \nonumber\\
  \textup{where}, & &   Q_{z_{1}}^{\textbf{r}_{0}}=\exp \left [ \frac{i \pi \left | \textbf{r}_{0}  \right |^{2}}{\lambda z_{1}} \right ] , \:   Q_{z_{1},z_{2}}^{\textbf{r}_{1},f}=\exp \left [ \frac{i \pi \left | \textbf{r}_{1}  \right |^{2}}{\lambda} \left ( \frac{1}{z_{1}}+\frac{1}{z_{2}}-\frac{1}{f}  \right )\right ],  \nonumber\\
& &  Q_{z_{2},f}^ {\textbf{r}_{2},\textbf{s}_{n}}=\exp \left [ \frac{i \pi}{\lambda } \left ( \frac{\left | \textbf{r}_{2}  \right |^{2}}{z_{2}}-\frac{\left | \textbf{s}_{n}  \right |^{2}}{f}  \right )\right ]  , \:  C_{z_{2}}^{z_{1}}=\frac{1 }{ \lambda^{2} z_{1}z_{2}}\exp \left [  \frac{i2 \pi (z_{1}+z_{2})}{\lambda } \right ].  \nonumber\\
\label{Eqn:DetectorAnalyticalExpression}
\end{eqnarray}

\noindent Equation~\ref{Eqn:DetectorAnalyticalExpression} is generally applicable to all systems in which a lens is placed between the object and detector. This includes two special cases:

\emph{The Fourier transforming condition}, where the object is placed very close to the lens plane ($z_{1}\rightarrow 0$) such that $\Psi_{in}= \Psi_{obj}$, the detector is placed in the focal plane ($z_{2}=f$), and the lens is completely transparent ($A_{\textbf{s}_{n}}^{\textbf{r}_{1}}=1$). In this configuration, the measured intensity of the ``focused field'' becomes the squared modulus of the Fourier transform of object field (i.e. $\left | \Psi_{det}  \right |^{2}\propto  \left | \mathcal{F} \left \{ \Psi_{obj} \right \} \right |^{2}$). This results in a variation of coherent diffraction imaging in which the lens can be used to reduce the large propagation distances necessary to achieve far-field diffraction patterns in the short wavelength regime \cite{Quiney2006}.

\emph{The imaging condition} - considered in this study - corresponds to where the object, lens and detector are placed according to the famous thin lens formula:
\begin{eqnarray} 
\frac{1}{f} = \frac{1}{z_{1}}+ \frac{1}{z_{2}}.
\label{Eqn:ThinLensFormula}
\end{eqnarray}
 
 \noindent At this condition the term $Q_{z_{1},z_{2}}^{\textbf{r}_{1},f}$ becomes unity, substantially simplifying Eq.~\ref{Eqn:DetectorAnalyticalExpression} in the context of the forward problem. More importantly, the detected image will resemble an inverted version of the object's exit surface ($I_{obj}$) - a valuable asset that will be exploited in the inverse problem described in section~\ref{InverseProblem}. 
 
\section{Methodology of lens translation imaging (LTI)} \label{Lens translation}

This section describes the image acquisition method for LTI. Returning our attention to the lens plane $\textbf{r}_{1}$ in Fig.~\ref{Fig:FourierPtychoSetup}, one sees that the finite size of the lens aperture means that a single $\left | \Psi_{det} \right |^{2}$ measurement will only register information corresponding to a limited region of $\left | \Psi_{in} \right |^{2}$. Therefore, acquiring multiple $\left | \Psi_{det} \right |^{2}$ measurement at different lens translation positions becomes paramount if one wishes to record a higher portion of spatial frequency data (real and complex) and subsequently improve the spatial resolution of the compound image. 

\begin{figure}[H]
\centering
\includegraphics[width=130mm]{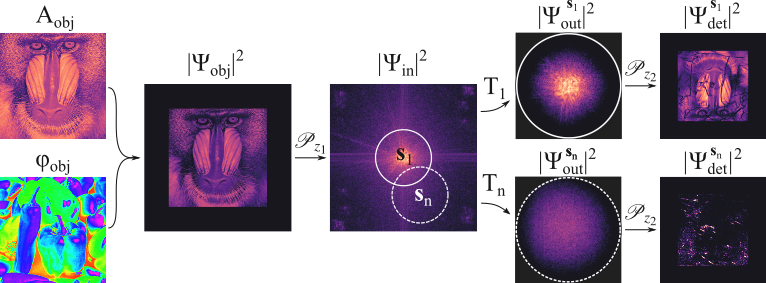}
\caption{Forward simulations and illustrating the LTI data acquisition process, from the input amplitude and phase (left), to the intensity at the lens plane (centre), to the resulting intensity on the detector (far right). Amplitude and intensity images are scaled from 0 to 1, while phase images are scaled from $-\pi$ to $+\pi$.} 
\label{Fig:LTI}
\end{figure}

Figure~\ref{Fig:LTI} depicts a flow-chart for the LTI methodology, in which a complex test image ($\Psi_{obj}$) is successively forward propagated to the lens plane ($\Psi_{in}$ and $\Psi_{out}$) and to the detector plane ($\Psi_{det}$). Importantly, the schematic embodies the key idea of LTI where the lens is translated to different position $\textbf{s}_{n}$ in a way that several overlapping areas of $\Psi_{in}$ are imaged at the detector. 

\noindent The forward simulations shown in Fig.~\ref{Fig:LTI} were chosen to be representative of a typical full-field transmission X-ray microscope operating at hard X-ray energies \cite{Snigireva2018} and with an X-ray magnification of approximately 18. The complex object wavefield (i.e. $\Psi_{obj}=A_{obj}e^{i \phi_{obj} }$, see far left of Fig.~\ref{Fig:LTI}) consisted of standard test images of a mandrill and peppers for the amplitude and phase, respectively. The physical size of this wavefield was 25.6 (W) $\times$ 25.6 (H) $\mu$m$^{2}$, and the wavelength was chosen to be $\lambda=0.75$ $\AA$, corresponding to photon a energy of 16.5 keV. This field is propagated by a distance $z_{1}=0.264$ m where the output field  $\left | \Psi_{in} \right |^{2}$ according to the Frensel number $N_{f}=\frac{\Delta x^{2}}{\lambda z_{1}}\approx  10^{-4}$ lies in the far-field regime for the given object pixel size of $\Delta x=50$ nm (not to be confused with the detector pixel size of $\approx$ 0.9 $\mu$m). For a focus distance of $f=0.25$ m the detector distance will be $z_{2}=4.736$ m to satisfy the condition in Eq.~\ref{Eqn:ThinLensFormula} yielding a geometrical magnification of M=17.9. To approximate experimental conditions, the detected intensity images incorporated Poisson noise, which varied from approximately 1.5 to 20 \% from the central to the outermost lens position. We note that the signal-to-noise is expected to decrease as the corresponding image intensity decreases towards the most distant lens positions. The attenuation properties of the lens were approximated by an apodized Gaussian distribution with a variance of $\sigma=25$ $\mu$m and a physical aperture of radius $r_{phys}=75 $ $\mu$m, which is typical for commercially-produced 2D Be-based compound refractive lenses (CRLs) with this focal length and energy \cite{Simons2017}. 

The far right of Fig.~\ref{Fig:LTI} shows two examples of simulated detected intensity images, labeled as corresponding to lens positions $\textbf{s}_{1}$ and $\textbf{s}_{n}$. As Eq.~\ref{Eqn:DetectorAnalyticalExpression} predicts, the intensity corresponding to the central axis position, $\left | \Psi^{\textbf{s}_{1}}_{det} \right |^{2}$ is approximately an inversion of $I_{obj}$. The resolution, however, is considerably poorer due to the masking of high spatial frequencies outside the aperture of $T_1$. Furthermore, residual features from sharp gradients in the phase map are visible as mild intensity variations in this centered image \cite{Zernike1942}. The off-centered image $\left | \Psi^{\textbf{s}_{n}}_{det} \right |^{2}$, corresponds to a region of $\Psi^{\textbf{s}_{1}}_{det}$ with mostly higher spatial frequency data. The intensity variations in this image thus reveal a higher fraction of morphological detail associated with the phase map, with visible features similar to those seen in differential inference contrast images \cite{Kaulich2002,Ou2013}. This type of contrast is typical in images attained using X-ray imaging techniques such as diffraction enhanced imaging \cite{Foster1980}, grating-based interferometry \cite{Pfeiffer2006}, and speckle-based phase-contrast \cite{Kaye2012}.      

\section{Iterative phase-retrieval (Inverse problem)} \label{InverseProblem}

\noindent The iterative phase-retrieval aims to recover the object wavefield ($\Psi_{obj}=A_{obj}e^{i \phi_{obj} }$) from a series of spatially overlapped translated images, each of which has only amplitude information.  

\begin{figure}[H]
\centering
\includegraphics[width=130mm]{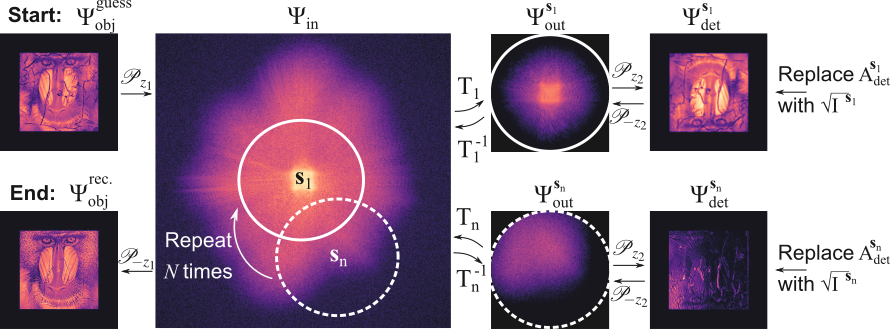}
\caption{ Flow-chart illustrating the iterative phase-retrieval procedure used in LTI.} 
\label{Fig:iterativeRecProcedure}
\end{figure}  

\noindent We explain the phase-retrieval procedure with the aid of Fig~\ref{Fig:iterativeRecProcedure}. While only two $ \textbf{s}_{n}$ positions are used for explanatory reasons, it can be trivially generalized to arbitrarily many positions ($n > 2 $). The procedure (described here in the case of $s_1$) is as follows: (1) make an initial guess of the object wavefield $\Psi^{guess}_{obj}$. (2) forward propagate $\Psi^{guess}_{obj}$ by a distance $z_{1}$ using Eq.~\ref{Eqn:QPF} to obtain $\Psi^{N=1}_{in}$. (3) multiply $\Psi^{N=1}_{in}$ by the lens transmission function for the position $ T_{1}$ to give $\Psi^{\textbf{s}_{1}}_{out}$. (4) forward propagate $\Psi^{\textbf{s}_{1}}_{out}$ by a distance $z_{2}$ to determine $\Psi^{\textbf{s}_{1}}_{det}$. (5) replace the amplitude with the square root of the measured intensity at that position. (6) back propagate by a distance $-z_{2}$ to give an updated $\Psi^{\textbf{s}_{1}}_{out}$. (7) Update $\Psi^{N=1}_{in}$ in the area corresponding to position $ \textbf{s}_{1}$ using the extended ptychographic iterative engine (E-PIE) approach of Rodenburg $et$ $al$.~\cite{Rodenburg2004}:
\begin{eqnarray} 
\Psi^{new}_{in} = \Psi^{old}_{in} + \frac{T_1^*}{\left | T_1 \right |^{2}_{max}}\left(\Psi^{\textbf{s}_{1}, new}_{out} - \Psi^{\textbf{s}_{1}, old}_{out}\right).
\label{Eqn:E-PIE}
\end{eqnarray}

\noindent (8) Move to the neighboring position $ \textbf{s}_{2}$ and repeat steps (3)-(7). This process (steps 1-8) is then carried out up to $ \textbf{s}_{n}$ and repeated for $N$ iterations or until the error metric $E_{0}$ has reached a minimum value. The error metric used here is defined as \cite{Maiden2009}:

\begin{eqnarray} 
E_{0}=  \frac{\sum _{\textbf{r}_{0}}  \left | \Psi_{obj}(\textbf{r}_{0})-\Psi^{N}_{rec}(\textbf{r}_{0}) \right |^{2} }{\sum _{\textbf{r}_{0}}  \left | \Psi_{obj}(\textbf{r}_{0})\right |^{2} }
\label{Eqn:ErrorMetric}
\end{eqnarray}

\noindent where, $\Psi^{N}_{rec}(\textbf{r}_{0})$ is the reconstructed object wavefield after a particular iteration $N$. The final step (9) involves back propagating the $N^{th}$ iteration of $\Psi^{N}_{in}$ by a distance $-z_{1}$, therefore fully recovering $\Psi^{rec}_{obj}$. 

\begin{figure}[H]
\centering
\includegraphics[width=130mm]{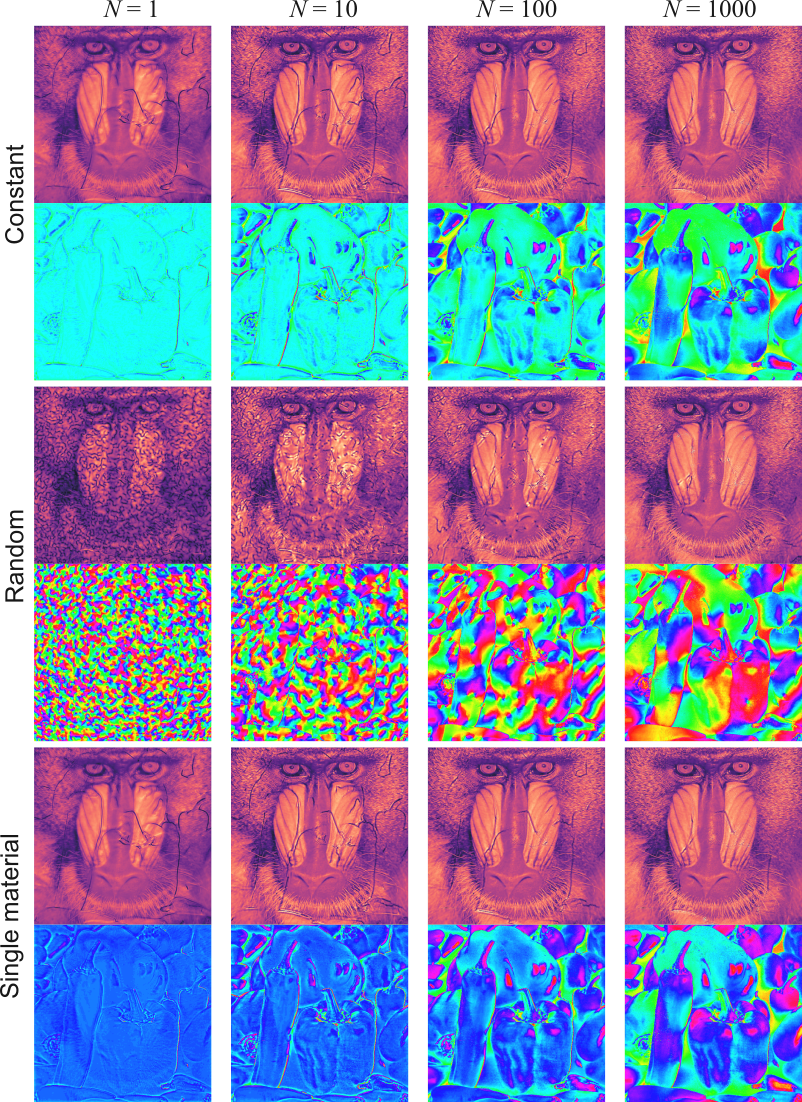}
\caption{Reconstruction of the object's wavefield performed with $N=1,\:10,\:100,$ and $1000$ iterations.} \label{Fig:RecResults}
\end{figure}

The reconstruction procedure was applied to a series of $n=25$ intensity images calculated for $n=25$ different lens positions corresponding to a 2.38$\times$ increase in NA with an average overlap of 80\% of their physical aperture (radius 150 $\mu$m). The initial guess utilized the centered intensity measurement as the initial guess of the object’s amplitude $|\Psi^{guess}_{obj}|=\sqrt{I^{\textbf{s}_{1}}}$. For the initial guess of the object's phase, three options were explored: (i) A phase grid with a constant value of $0$ across the plane (i.e. $\Psi^{guess}_{obj}=\sqrt{I_{1}} e^{i\phi_{constant}}$); (ii) A phase grid generated with statistically random values that fluctuate uniformly between $[0,2 \pi]$ ($\Psi^{guess}_{obj}=\sqrt{I_{1}} e^{i\phi_{random}}$); (iii) A phase grid constructed using the relation between intensity and phase based on the arguments made by Paganin $et$ $al$.~\cite{Paganin2002}. That is, 

\begin{eqnarray} 
\phi_{guess}= \frac{\gamma}{2} \ln (I^{\textbf{s}_{1}}),
\label{Eqn:SingleMaterial}
\end{eqnarray}

\noindent where the ratio $\gamma=\frac{\delta}{\beta}$ relates to the object's complex refractive index distribution $n_{r}=1-\delta+i\beta$. A key assumption of this relation is that the value of $\gamma$ is constant throughout the object's volume predicating it is largely composed of a single-material. To test the effectiveness of each guess, the reconstruction of $\Psi^{rec}_{obj}$ was attained after $N=1, 10,\:100,$ and $1000$. The respective results are shown in Fig.~\ref{Fig:RecResults}. 

\noindent In all three cases, the amplitude reconstructions presented in Fig.~\ref{Fig:RecResults} require fewer iterations to reach an acceptable solution in comparison to the phase. This is primarily due to the understandably close resemblance of the initial guess to the true amplitude of the exit wavefield at the object. We additionally note that the speckle-like noise pollution for the random phase guess below $N=100$ is likely due to the strong phase gradients being manifested in the intensity, and disappears by $N=1000$.

\begin{figure}[H]
\centering
\includegraphics[width=90mm]{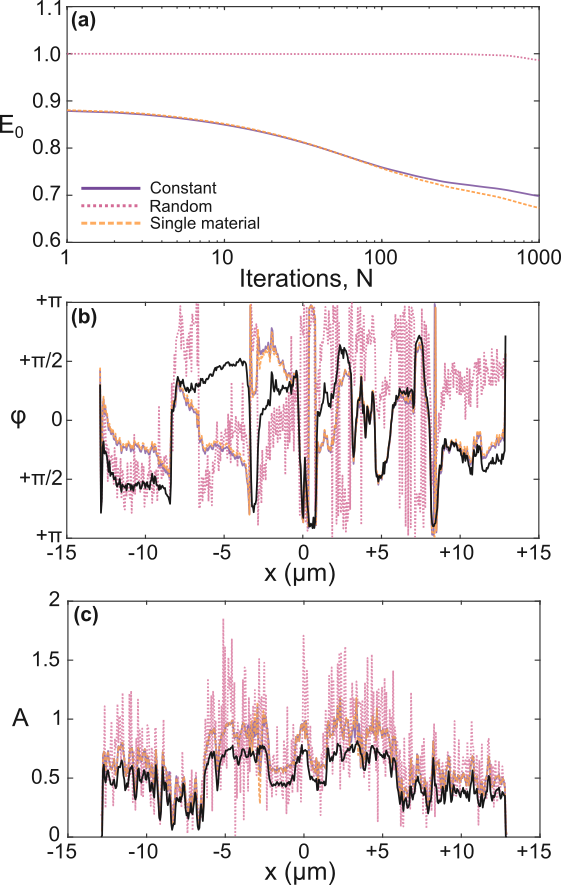}
\caption{(a) Plot showing the evolution of $E_{0}$ versus number of iterations $N$ for all initial phase guess choices. (b) and (c) display overlaid profiles taken horizontally across the centre of the reconstructed phase and amplitude images for $N=1000$. The black curve corresponds to the original (true) phase and amplitude input map.} 
\label{Fig:Plots}
\end{figure} 

\noindent The choice of $\phi_{guess}$ clearly has a decisive effect on the quality and convergence rate of the final phase reconstructions, shown in Fig.~\ref{Fig:Plots}. Both the convergence plot (Fig.~\ref{Fig:Plots}a) an the quantitative accuracy of the amplitude and phase (Fig.~\ref{Fig:Plots}b,c) strongly favour the flat phase or single material assumption over the random phase guess. This is a somewhat surprising observation, as the test phase image (Peppers) contained large variations over $2\pi$, including significant phase gradients. These large phase gradients appeared to cause some significant errors in the recovered phase related to phase-wrapping, though in general the recovered phase is quantitatively similar to the phase of the original test image. More surprising, however, is the improved convergence rate of the single material assumption from $N>200$, given that there was no correlation between the phase and amplitude of the test image, which undoubtedly violates its key premise in Eq.~\ref{Eqn:SingleMaterial}. This observation supports recent work by Gureyev $et$ $al$.~\cite{Gureyev2015}, which showed that the single material assumption can extend to a broader class of samples without significant loss of generality. For further details, the reader is encouraged to refer to Ref. ~\cite{Gureyev2015}.

\section{Discussion and conclusion}

Lens translation imaging (LTI) provides a practical approach to synthetically increasing the numerical aperture, spatial bandwidth product and phase sensitivity of classical full-field hard X-ray microscopes. The methodology is described here in detail using coherent scalar wave optics theory to derive a generalized mathematical expression for the wavefield as it traverses the entire LTI system, and includes a formulation of the iterative phase-retrieval algorithm based on the popular E-PIE algorithm. 

In addition to providing the mathematical framework for developing simulation and image recovery code, the analytical expressions also provide valuable physical insights into the image contrast mechanism. In particular, we note that the forward simulations (based on Eq.~\ref{Eqn:DetectorAnalyticalExpression}) suggest that the off-axis intensity images contain clear contrast in the form of differential phase contrast (DPC). By paying specific attention to the lens amplitude function $A_{\textbf{s}_{n}}^{\textbf{r}_{1}}$, we note the term $\exp\left [ (\textbf{s}_{n}\cdot\textbf{r}_{1})/\sigma^{2} \right ]$ arises once the squared binomial $\left | \textbf{r}_{1}- \textbf{s}_{n} \right |^{2}$ is expanded. Taylor approximating this term to the first order then invoking the Fourier derivative theorem explains the origin of the DPC signal and how its contribution is proportional to the shifting $\left | \textbf{s}_{n} \right |$. From this it becomes clear that this type of contrast is the same as that observed and studied in visible light FPM setups \cite{Ou2013}.    

Realizing LTI means that the compound image must be recovered without the access to the true amplitude and phase maps as benchmarks for convergence, as was the case here in Eq.~\ref{Eqn:ErrorMetric}. To this end, we recommend to calculate the error metric relative to the measure images taken at the various lens positions via the following formula:

\begin{eqnarray} 
E_{0}= \frac{1}{n}\sum _{n} \left ( \frac{\sum _{\textbf{r}_{2}}  \left | \sqrt{I^{\textbf{s}_{n}}_{meas}(\textbf{r}_{2})}-\sqrt{I^{\textbf{s}_{n}}_{N}(\textbf{r}_{2})} \right |^{2} }{\sum _{\textbf{r}_{2}}  I^{\textbf{s}_{n}}_{meas}(\textbf{r}_{2})}   \right ),
\label{Eqn:ErrorMetricExperiment}
\end{eqnarray}

\noindent where, $I^{\textbf{s}_{n}}_{meas}(\textbf{r}_{2})$ is the measured intensity for a certain $\textbf{s}_{n} $ and $\sqrt{I^{\textbf{s}_{n}}_{N}(\textbf{r}_{2})}$ is the calculated intensity at the same $\textbf{s}_{n}$ for a particular iteration $N$. The above error metric takes into consideration the average over all lens positions.

Our study of the iterative phase recovery revealed the importance of the initial guess on the rate of convergence and the ultimate quality of the compound image. Like FPM, LTI has an extremely significant advantage that the central image may be used as an accurate guess for the amplitude. However, although the single-material assumption offers some advantages of a constant phase guess, we believe there is still room for improvement. Given that elements of DPC are clearly present in the off-axis images, it seems intuitive that this information could be utilized to provide a more accurate guess of the object phase that would significantly improve the convergence rate. 

As a full-field X-ray imaging technique, LTI might have the potential to offers a significantly reduced radiation dose rate (as opposed to accumulated dose) in comparison to scanning-probe methods, such as conventional X-ray ptychography. While the mechanisms of radiation damage vary greatly between specimens, many have clear dependencies on the radiation dose rate \cite{berejnov2018}. In the typical cases of a scanning nanoprobe with a 200 nm probe diameter and a full-field microscope with a 200 $\mu$m diameter, one can anticipate a reduction in dose rate of 6 orders of magnitude. In the case of the latest generation of high-brilliance coherent X-ray source \cite{Eriksson2014}, this reduction may be key. 

Perhaps most importantly, however, LTI has the potential to be both intuitive and widely accessible. By retaining the ``what you see is what you get'' character of full-field microscopy, it provides users the ability to quickly and decisively design and perform measurements, and to seamlessly switch from fast overviews of the entire specimen (narrow scan range) to detailed inspections of individual elements (broad scan range). Because LTI is based on the classical Galilean geometry used by full-field microscopes around the world, it can also be implemented with little-to-no additional hardware. Given the imminent improvements in brilliance and coherence of synchrotron sources, we believe LTI could be a convenient, valuable and effective new tool for the broad spectrum of X-ray microscopists. 

\section*{Acknowledgements}
The authors received financial support from a VILLUM Experiment grant and ERC Starting Grant (3D-PXM). In addition, they gratefully acknowledge helpful discussions with D. Paganin, H.F. Poulsen and A.F. Pedersen. 


\end{document}